\documentclass[preprint,12pt,sort&compress]{elsarticle}

\usepackage{amsmath,amssymb,mathrsfs}
\usepackage[]{graphicx}
\usepackage[latin1]{inputenc}
\usepackage{color}
\usepackage{enumerate}
\usepackage{hyperref}
\usepackage{caption}
\hypersetup{colorlinks=true,citecolor=blue}

\journal{Chemical Engineering Science}

\begin{document}

\begin{frontmatter}

\title{Modelling of artefacts in estimations of particle size of needle-like particles from laser diffraction measurements}

\author[a1]{Okpeafoh S. Agimelen\corref{cor1}}
\ead{okpeafoh.agimelen@strath.ac.uk}

\author[a2]{Anthony J. Mulholland}

\author[a1]{Jan Sefcik\corref{cor1}}
\ead{jan.sefcik@strath.ac.uk}

\cortext[cor1]{Corresponding authors}
\address[a1]{EPSRC Centre for Innovative Manufacturing in Continuous Manufacturing and Crystallisation, Department of Chemical and Process Engineering, University of Strathclyde, James Weir Building, 75 Montrose Street, Glasgow, G1 1XJ, United Kingdom.}

\address[a2]{Department of Mathematics and Statistics, University of Strathclyde, Livingstone Tower, 26 Richmond Street, Glasgow G1 1XH, United Kingdom.}

\begin{abstract}

Manufacturing of particulate products across many industries relies on accurate measurements of particle size distributions in dispersions or powders. Laser diffraction (or small angle light scattering) is commonly used, usually off-line, for particle size measurements. The estimation of particle sizes by this method requires the solution of an inverse problem using a suitable scattering model that takes into account size, shape and optical properties of the particles. However, laser diffraction instruments are usually accompanied by software that employs a default scattering model for spherical particles, which is then used to solve the inverse problem even though a significant number of particulate products occur in strongly non-spherical shapes such as needles. In this work, we demonstrate that using the spherical model for the estimation of sizes of needle-like particles can lead to the appearance of artefacts in the form of multimodal populations of particles with size modes much smaller than those actually present in the sample.
This effect can result in a significant under-estimation of the mean particle size and in false modes in estimated particles size distributions.

\end{abstract}


\end{frontmatter}

\section{Introduction}
\label{sec1}

Particle size measurements are crucial across many industries in the manufacturing of particulate products, such as pharmaceuticals, agrochemicals, detergents, pigments and food. Particle size and shape have a profound influence on downstream processing as well as on final product properties through a variety of product attributes, such as solubility, dissolution kinetics, flowability, etc. There are various particles sizing methods commonly employed in practice, some of them online and others offline \cite{Washington1992,Shekunov2007,Abbireddy2009}. One of the widely used techniques for measuring of particle size distribution (PSD) in dispersions is laser diffraction \cite{Black1996} which is typically arranged in a flow-through setting but is most often used off line as there are limits on dispersion densities due to multiple scattering.
Laser diffraction measurement involves the collection of scattered light from a dilute dispersion of particles by an array of detectors placed at different spatial locations so that they cover a certain span of scattering angles $\theta$. Since the angular dependence of the scattered light intensity originating from a particle is a function of the size and shape of the particle, as well as the orientation of the particle with respect to the incident laser beam, the particle size and shape can be inferred from the corresponding scattering intensity pattern. However, as there is typically a distribution of particle sizes across a population, the intensity pattern measured by the detectors will be a convolution of the intensity patterns from all the particles of different sizes in the dispersion. 

The estimation of the PSD from the measured scattering intensity pattern (scattering intensity as a function of scattering angle) involves solving an inverse problem using a suitable scattering model which describes the scattering intensity for particles of a given shape, size and optical properties. The inversion is implemented in the software accompanying laser diffraction instruments, typically using the Mie scattering model \cite{Bohren1983} for spherical particles as a default, regardless of the shape of particles in the measured sample. This can lead to various artefacts, such as apparently multimodal distributions in PSD estimates (e.g., \cite{Hamilton2012,Polakowski2014}) when the shape of the particles in the sample deviates significantly from spherical. This could result in misleading estimates of mean particles sizes with severe consequences for applications where the process is very sensitive to the particle sizes. This is particularly important in the pharmaceutical industry where many of the active pharmaceutical ingredients are crystallised in needle-like habits. 

In this paper, we demonstrate that multimodal PSDs can be obtained from the inversion process even when the true particle size is monodisperse, i.e., all particles are of the same size.  We simulate the scattering intensity patterns for monodisperse population of needle-like particles to explain how multimodal PSD artefacts arise due to solving the inverse problem using a scattering model for spherical particles. We compute the angularly dependent scattering intensity for needle-like particles of specified optical properties using a model for infinitely long cylinders with diameters ranging from 1 to 100 micrometres.  Then we solve the inverse problem of estimating PSD from the angularly dependent scattering intensity pattern using the Mie theory for spherical particles, mimicking the analysis performed when using commercial laser diffraction instruments. Since the PSD of needle-like particles is exactly specified here, the estimated PSD from the inversion can be directly compared with the actual PSD.
We consider two limiting cases, when needle-like particles are assumed to perfectly aligned with flow which is perpendicular to the incident laser beam and when they adopt random orientations. We show that in either case the inversion results in estimated PSD that are multimodal where smaller modes are mathematical artefacts due to inversion and we explain how these are related to different shapes of intensity scattering patterns for needle-like particles compared to those of spheres.

\section{Calculating scattering intensity}
\label{sec2}

\begin{figure}[tbh]
\centerline{\includegraphics[width=\textwidth]{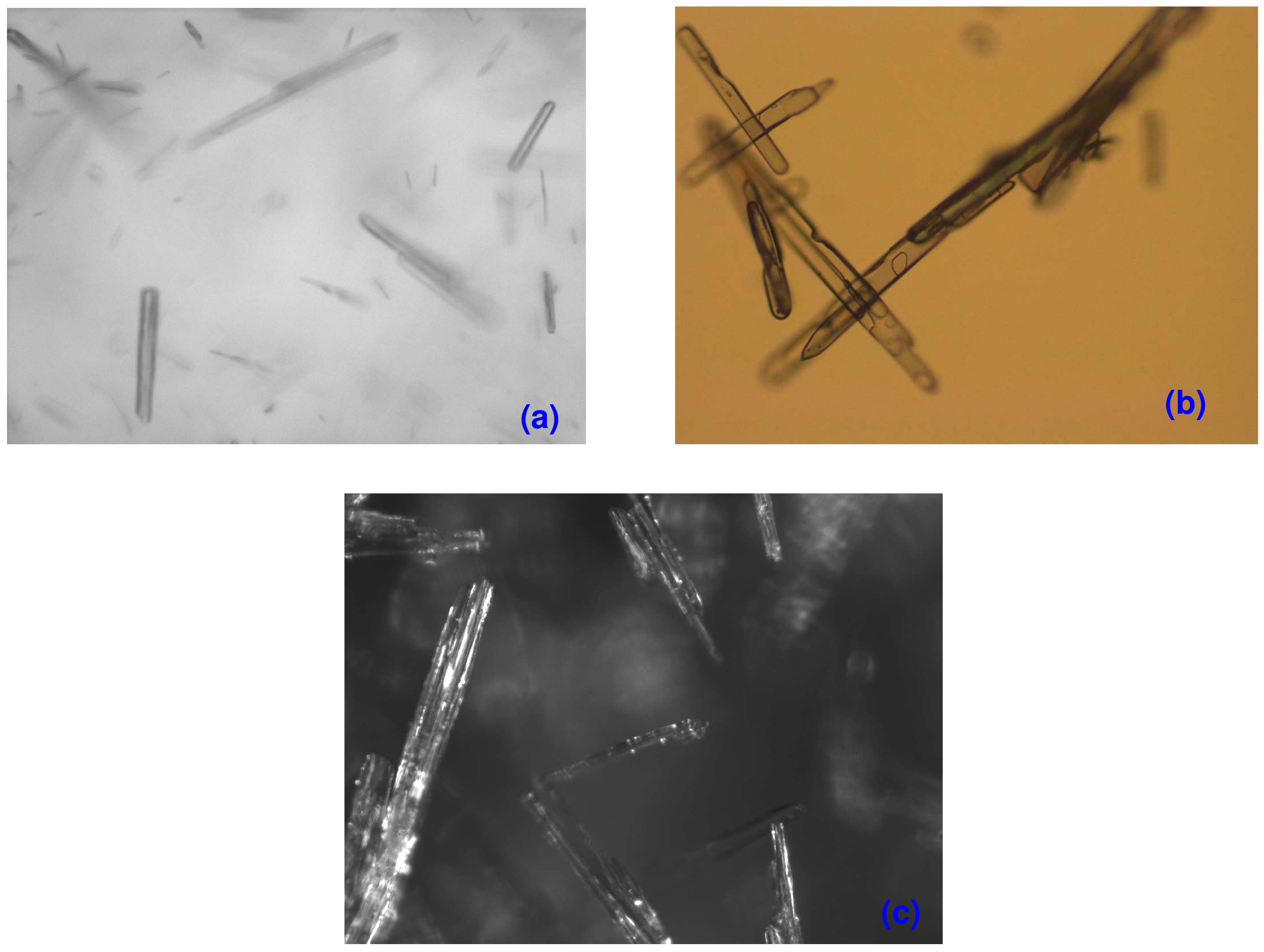}}
\caption{Images of typical needle-like crystals of (a)\,cellobiose octaacetate, (b)\,benzoic acid and (c)\, metformin hydrochloride.}
\label{fig1}
 \end{figure}

The scattering intensity patterns for needle-like particles will be simulated using the scattering theory for infinitely long cylinders \cite{Bohren1983}. Even though the theory was developed for infinitely long cylinders, it is applicable for needle-like particles with lengths significantly larger than their diameters \cite{Wickramasinghe1973}. Such long thin particles of approximately cylindrical shape are similar to needle-like particles often encountered in pharmaceutical and chemical manufacturing. For example, in Fig. \ref{fig1} we show typical particles of cellobiose octaacetate, benzoic acid and metformin hydrochloride.
As needle-like particles are modelled as infinitely long cylinders in this work, the circular cross-sectional diameters of these cylinders will be used to represent the size of needle-like particles as their length cannot be specified.
 The scattering intensity pattern for spherical particles will be simulated using the Mie theory (see \cite{Bohren1983} and section 1 of the supplementary information for details). The procedure for performing the calculations is described below.

\begin{figure}[tbh]
\centerline{\includegraphics[width=\textwidth]{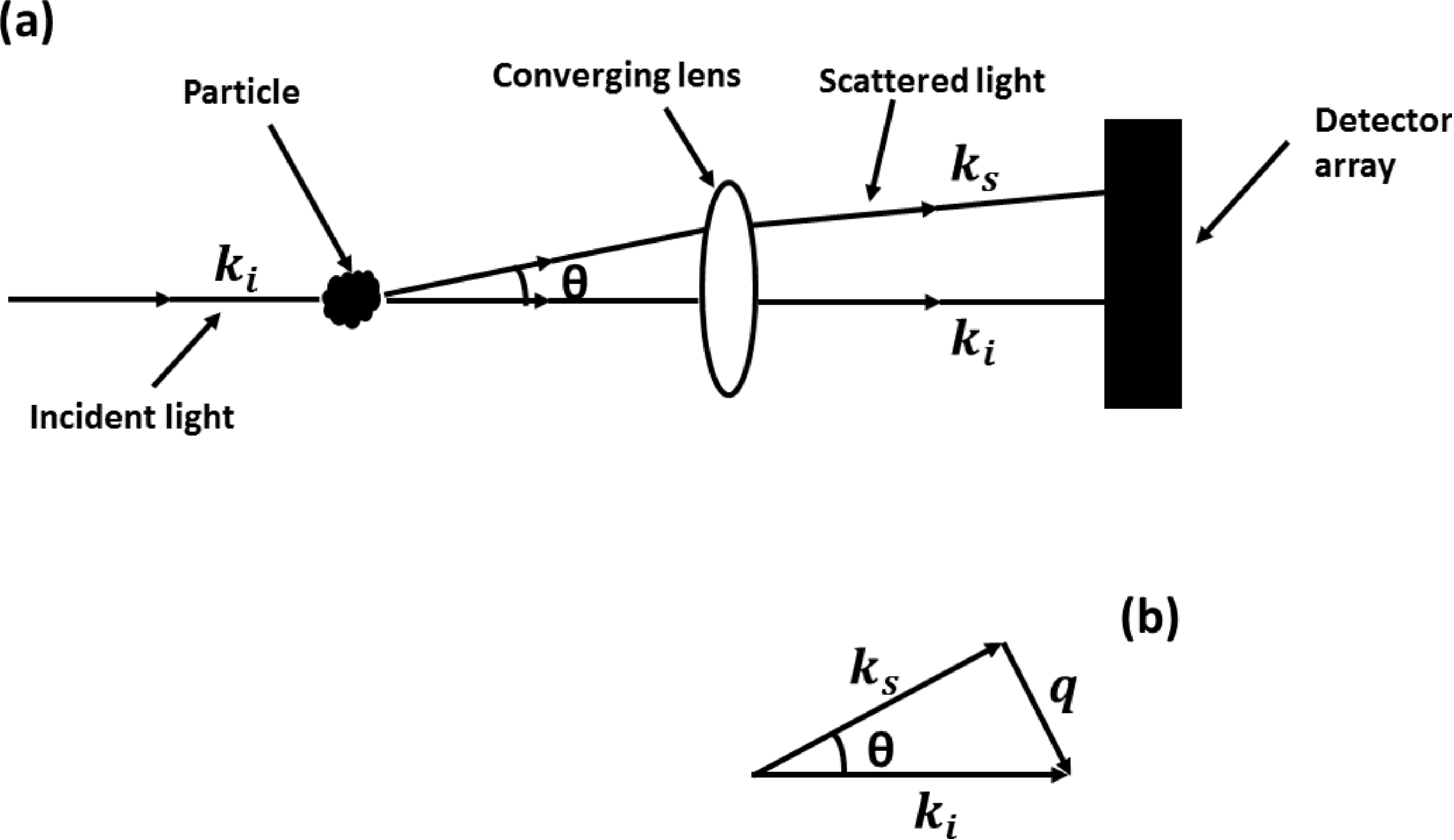}}
\caption{(a)\,Schematic of the setup of typical laser diffraction instruments. (b)\,Illustration of the scattering wave vector.}
\label{fig2}
 \end{figure}

Consider a detector system (sketched in Fig. \ref{fig2}(a)) in which a monochromatic light  with wave vector $\mathbf{k}_i$ is incident on a particle of arbitrary size and shape. The scattered light with wave vector $\mathbf{k}_s$ is then collected at different angles $\theta$ to the direction of propagation of the incident light by an array of detectors as depicted in Fig. \ref{fig2}(a). Both the incident and scattered light have components parallel and perpendicular to the scattering plane (the plane containing the incident and scattered light) \cite{Bohren1983}. The scattering wave vector $\mathbf{q}$ is the difference between the incident and scattered wave vectors as sketched in Fig. \ref{fig2}(b). The magnitude of the scattering wave vector is a function of the scattering angle $\theta$, and it is given by \cite{Sorensen2001}

\begin{equation}
q = \frac{4\pi}{\lambda}\sin(\theta/2),
\label{eq1}
\end{equation}
where $\lambda$ is the wavelength of the incident light. The wavelength of $\lambda = 0.633\mu$m was used in all calculations in this work consistent with the wavelength of red light in typical commercial laser diffraction instruments.

The intensity of the scattered light $\mathbf{I}$ (which is a function of the scattering angle $\theta$) is related to the intensity of the incident light $\mathbf{I}_0$ by means of the phase or scattering matrix $\mathbf{Z}$ as \cite{Bohren1983,Mishchenko2000} (see also section 1 of the supplementary information for details)

\begin{equation}
I(q_j) = \frac{I_0}{k^2r^2}\sum_{i=1}^{N}Z_{11}(q_j,\overline{D}_i)X(\overline{D}_i).
\label{eq2}
\end{equation}
where the magnitude of the wave vector $q$ has been discretised\footnote{The values of scattering angles $\theta = 0.0015^{\circ}$ to $\theta = 180^{\circ}$ were used in all the calculations in this work. This is to ensure that the entire forward direction for scattering is covered. The quantity $1/q$ has the dimension of length, and it sets the length scale accessible in a light scattering experiment \cite{Sorensen2001}. The smallest and largest values of $\theta$ correspond to length dimensions $1/q$ of order $3800\mu$m and $0.05\mu$m respectively. These are the same order of magnitude as the largest and smallest particle diameters of $D_{N+1}=3500\mu$m and $D_1=0.01\mu$m (respectively) in the particle size grid used in the calculations. The angular positions $\theta$ were discretised on a geometrically spaced grid with 100 bins. This then gives a size difference of order $L^2\Delta q\,\mu$m (where $L$ is an arbitrary particle size and $\Delta q$ is the $q$ spacing). As low $q$ ($\Delta q \approx 10^{-4}\mu\textrm{m}^{-1}$) values correspond to large particle sizes, then using $L=3800\mu$m gives a size difference of $144\mu$m, which is about 4\% of $D_{N+1}$. This is the resolution of the largest particle size. In the large $q$ ($\Delta q \approx 1\mu\textrm{m}^{-1}$) region, which corresponds to small particles, the size difference is of order $10^{-4}\mu$m (using $L=0.01\mu$m) which is about 1\% of $D_1$, and this is the resolution of the smallest particle size.}
into $j=1,2\ldots,M$ angular positions. The particle size (represented by diameter of spherical or circular cross section of needle-like particles modelled as infinitely long cylinders in this work) $D_i,\, i=1,2,\ldots,N$ has been discretised into $N$ particle size classes\footnote{A particle size grid with $N=200$ size classes was used in the calculations here. The particle diameters on the grid run from $D_1=0.01\mu$m to $D_{N+1}=3500\mu$m. This covers the entire range of particle diameters in typical commercial laser diffraction instruments \cite{Mastersizer3000}.}
 such that the characteristic diameter $\overline{D}_i$ of particles whose diameters lie between $D_i$ and $D_{i+1}$ is given by $\overline{D}_i=\sqrt{D_iD_{i+1}}$. The number of particles whose diameters lie between $D_i$ and $D_{i+1}$ is given by $X(\overline{D}_i)$. Finally, the quantity $Z_{11}(q_j,\overline{D}_i)$ is the first component (see section 1 of the supplementary information) of the phase matrix averaged over particles whose sizes lie between $D_i$ and $D_{i+1}$.

Writing $I(q_j)$ as $I_j$, $Z_{11}(q_j,\overline{D}_i)$ as $\tilde{Z}_{j,i}$ and $X(\overline{D}_i)$ as $X_i$, then Eq. \eqref{eq2} can be written as a matrix equation as
\begin{equation}
I_j = \mathbf{I} = \frac{I_0}{k^2r^2}\sum_{i=1}^{N}\tilde{Z}_{j,i}X_i = \frac{I_0}{k^2r^2}\mathbf{\tilde{Z}X}.
\label{eq3}
\end{equation}

The scaling factor in Eq. \eqref{eq3} can be removed if the scattering intensity is rescaled by the scattering intensity $I_1$ measured at an angular position close to zero. That is, the zero $q$ limit of the scattering intensity. The choice of $I_1$ for rescaling the scattering intensity is reasonable since the quantity $Z_{11}$ assumes a flat profile at the zero $q$ limit for particles of different shapes and sizes and refractive indices \cite{Bohren1983}. If the first detector in the array of detectors (sketched in Fig. \ref{fig1}) is placed at an angular position sufficiently close to zero, then the scattering intensity at the zero $q$ limit for the particles in the $N$ size classes can be constructed from the first row of matrix $\mathbf{\tilde{Z}}$ as
\begin{equation}
I_1 = \frac{I_0}{k^2r^2}\sum_{i=1}^{N}\tilde{Z}_{1,i}X_i.
\label{eq4}
\end{equation}
Then the rescaled scattering intensity $\mathbf{\tilde{I}}_j$ can be constructed as
\begin{equation}
\tilde{I}_j = \frac{I_j}{I_1} = \frac{\sum_{i=1}^{N}\tilde{Z}_{j,i}X_i}{\sum_{i=1}^{N}\tilde{Z}_{1,i}X_i}. 
\label{eq5}
\end{equation}
The rescaled scattering intensity in Eq. \eqref{eq5} is the form in which the scattering intensity data is reported in typical commercial laser diffraction instruments. Hence this form of the scattering intensity will be used in subsequent analysis in this work. 
The components of the phase matrix take different forms depending on the shape of the particles (see section 1 of the supplementary information).

\section{Forward  and inverse problem}
\label{sec3}

In a dispersion of particles of different sizes, the scattering intensity $\tilde{\mathbf{I}}^{\ast}$ that is measured   will be a convolution of the scattering intensities of the individual particles in the dispersion and the PSD $\mathbf{X}$ given by Eq. \eqref{eq2}. If the PSD $\mathbf{X}$ of the particles in the dispersion is given, then the scattering intensity of the population can be calculated by solving the forward problem in Eq. \eqref{eq5}. The calculated scattering intensity $\mathbf{\tilde{I}}$ can then be compared with the measured scattering intensity.

In reality the PSD $\mathbf{X}$ of the particles in a dispersion will not be known. Instead the challenge will be to estimate the PSD corresponding to a measured scattering intensity. This will involve solving an inverse problem. As mentioned earlier, the purpose of this work is to examine the effect of the model employed in solving the inverse problem on the solutions obtained. To achieve this objective, the experimentally measured scattering intensity $\mathbf{\tilde{I}}^{\ast}$ will be simulated with the model for infinitely long cylinders. The scattering intensity $\mathbf{\tilde{I}}$ calculated using the model for spherical particles 
(see section 1 of the supplementary information for details)
 will then be used to fit a given simulated intensity $\mathbf{\tilde{I}}^{\ast}$.

In this work, the inversion
required
to obtain the PSD $\mathbf{X}$ will be carried out by 
minimising the objective function $f$ given as
\begin{equation}
 f = \sum_{j=1}^M{w_j\left[\tilde{I}_j^{\ast} -  \tilde{I}_j\right]^2},
\label{eq6} 
\end{equation}
This is a weighted least square problem which is an unconstrained optimisation problem \cite{Boyd2004}. The weight function $w_j$ is given as
\begin{equation}
w_j = \frac{1}{1 + |C_1|\tilde{I}_j + |C_2|\tilde{I}_j^2},
\label{eq7} 
\end{equation}
where the quantities $C_1$ and $C_2$ are optimisation parameters with initial values $C_1 = C_2 = 0$. The weighting function in the objective function in Eq. \eqref{eq6} is necessary as the values of the scattering intensity cover several orders of magnitude over the entire $q$ range of interest. A similar weighting function was employed previously \cite{Schenk1998} for the calculation of intensity for anti-Stokes Raman scattering but with fixed values of $C_1$ and $C_2$. 

\subsection{Number and volume based PSD}
\label{ssec3-1}

The PSD defined in Eq. \eqref{eq5} which is calculated by solving the least square problem in Eq. \eqref{eq6} is number based. The number based PSD $X_i$ is defined as an exponential function of the parameter $\gamma_i$ as\footnote{The formulation of the PSD as an exponential function with the free parameter $\gamma$ ensures that the PSD is non negative and free to assume any shape. A prescribed shape for the PSD would bias the estimate as the shape of the PSD would have limited degrees of freedom.}
\begin{equation}
X_i = e^{\gamma_i}, i=1,2,\ldots,N.
\label{eq8}
\end{equation}
Then the optimisation problem given in Eq. \eqref{eq6} is solved by searching for $\gamma_i$ (using the Levenberg-Marquardt algorithm as implemented in Matlab) which minimises the objective function $f$ given in Eq. \eqref{eq6}. As the Levenberg-Marquardt method is gradient based and the objective function in Eq. \eqref{eq6} contains local minima, then a multi-start strategy \cite{Aster2013} is used to search for a  solution which is close to the global minimum. Once a solution close to the global minimum is found, then it is used to estimate the volume based PSD $X_i^v$ as described in section 2 of the supplementary information. Finally, the volume based PSD is normalised as 
\begin{equation}
\hat{X}^v_i = \frac{X^v_i}{\sum_{i=1}^{N}{X^v_i}}
\label{eq9}
\end{equation}
where $\hat{X}^v_i$ is the volume fraction of spherical particles of size $\overline{D}_i$.

\subsection{Mean particle size}
\label{ssec3-2}

The mean size of the particles in a population can be represented by various metrics depending on the application \cite{Merkus2009}. The volume weighted mean diameter $D_{43}$ is commonly reported by commercial laser diffraction instruments. The $D_{43}$ value is defined as \cite{Merkus2009}
\begin{equation}
D_{43} = \frac{\sum_{i=1}^{N}{X_i\overline{D}_i^4}}{\sum_{i=1}^{N}{X_i\overline{D}_i^3}},
\label{eq10}
\end{equation}
which upon using the substitution $X^v_i = X_iv_i$ (where $v_i$ is the volume of the spherical particle with diameter $\overline{D}_i$) becomes
\begin{equation}
D_{43} = \sum_{i=1}^{N}{\hat{X}^v_i\overline{D}_i},
\label{eq11}
\end{equation}
 The $D_{43}$ value will coincide with the diameter of spherical particles in a monodispersed population. 

\section{Results and Discussion}
\label{sec4}

Suspended needle-like particles experience various kinds of flow conditions depending on the vessel shape, agitation arrangement, solid loading and so on. Depending on these conditions, the particles may be able to perform completely random or partially restricted rotations or they could become fully aligned with the flow field. Flow-through cells typically used in laser diffraction instruments are a few millimetres wide where the incident laser beam is perpendicular to the flat transparent windows enclosing the cell and to the direction of flow through the cell. Depending on the shape, size and solid loading of particles, their rotations in such flow through cells may be restricted and in the limit of thin needles or thin platelets at high solid loadings they can become fully aligned with the flow with their main axis being perpendicular to the incident laser beam. Motivated by this, we consider two limiting cases: case I, where the incident beam is perpendicular to the main axis of the particle, and case II, where the incident beam is at a random angle to the main axis of the particle. These two cases will be examined in detail below.

 When the particles are free to perform random rotations, 
 the incident angle of the incoming monochromatic light will take all possible values from grazing to normal incidence
 with equal probability. However, when the particles are aligned with the flow field, the incident laser beam will be perpendicular to the particle main axis and hence the incident angle is fixed at 90$^{\circ}$ 
 (see Fig. 1 of the supplementary information for a schematic for light scattering by an infinitely long cylinder.).
  \begin{figure}
  \centerline{\captionsetup{width=\textwidth}
  \includegraphics[width=0.7\textwidth]{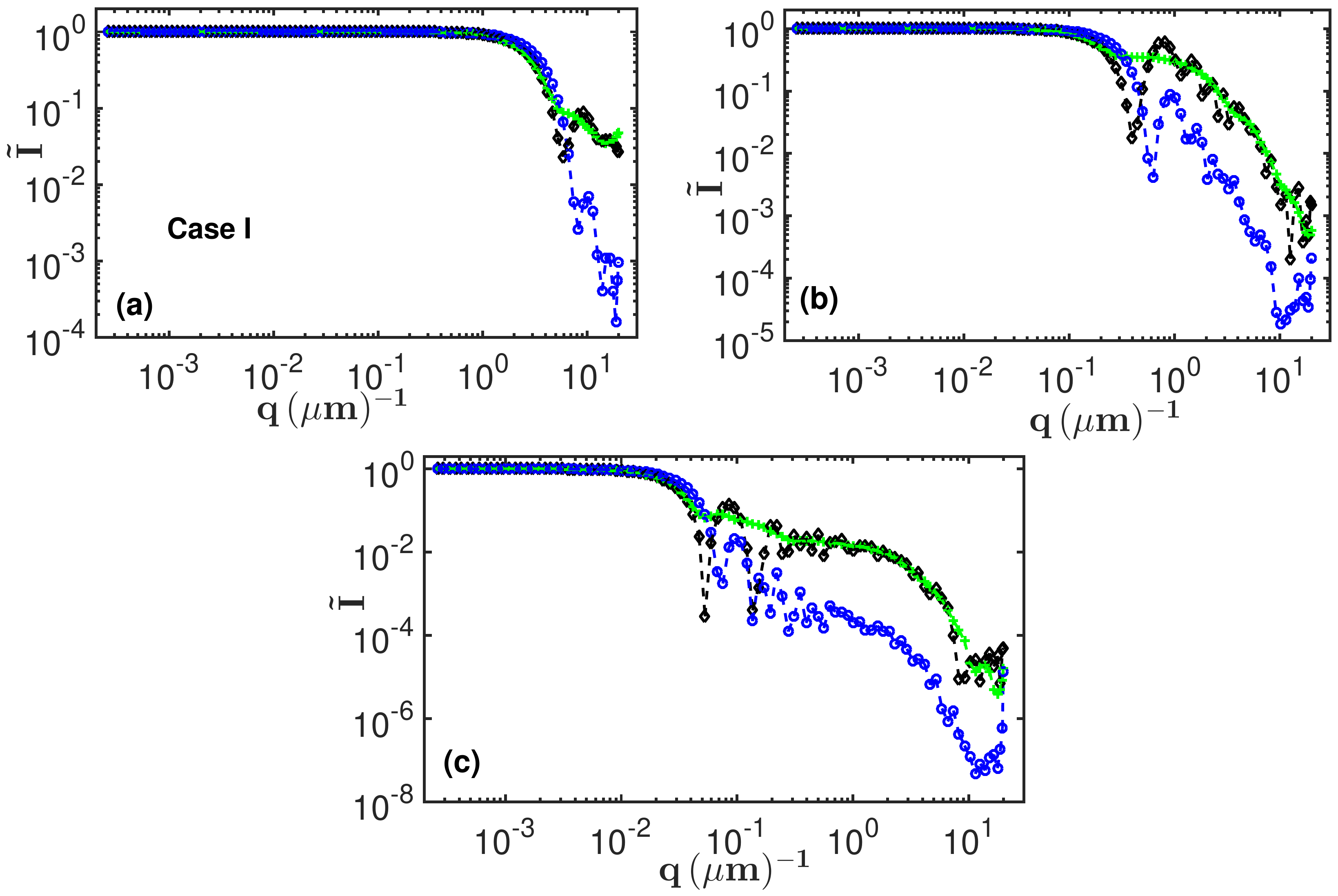}}
  \caption{
  Comparison of normalised scattering intensity patterns for needle-like and spherical particles of various sizes. Black diamonds: calculated scattering intensities for monodisperse needle-like particles with circular cross-section diameter of (a)\,$\overline{D}=1\mu$m, (b)\,$\overline{D}=10\mu$m and (c)\,$\overline{D}=100\mu$m for Case I. Green crosses: best fit (using the scattering model for a multimodal distribution of spherical particles) to the scattering intensity of monodisperse needle-like particles obtained by solving the optimisation problem in Eq. \eqref{eq6}. Blue circles: calculated scattering intensities for monodisperse spherical particles with diameter of (a)\,$\overline{D}=1\mu$m, (b)\,$\overline{D}=10\mu$m and (c)\,$\overline{D}=100\mu$m. The needle-like particles were modelled as infinitely long cylinders
  \label{fig3}}
   \end{figure}

The refractive indices of materials typically encountered in organic crystals (e.g., in pharmaceutical manufacturing) are around $N_r=1.50$ with zero absorption. For example, the cellobiose octaacetate, benzoic acid and metformin hydrochloride crystals shown in Fig. \ref{fig1} have refractive indices of $N_r = 1.51, 1.50$ and 1.58 respectively. The three materials have poor solubility in methanol which has a refractive index $N_r=1.33$. Hence the refractive index of $N_r=1.50$ was used in the simulation of the scattering intensities of needle-like particles (modelled as infinitely long cylinders), and the particles were assumed to be suspended in a non-solvent medium with refractive index $N_r=1.33$.  The wavelength of the incident light was fixed at $\lambda = 0.633\mu$m in all the calculations which is consistent with the wavelength of red light in typical commercial laser diffraction instruments.

\subsection{Case I: restricted rotations}
\label{ssec4-1}

 \begin{figure}[tbh]
\centerline{\includegraphics[width=0.8\textwidth]{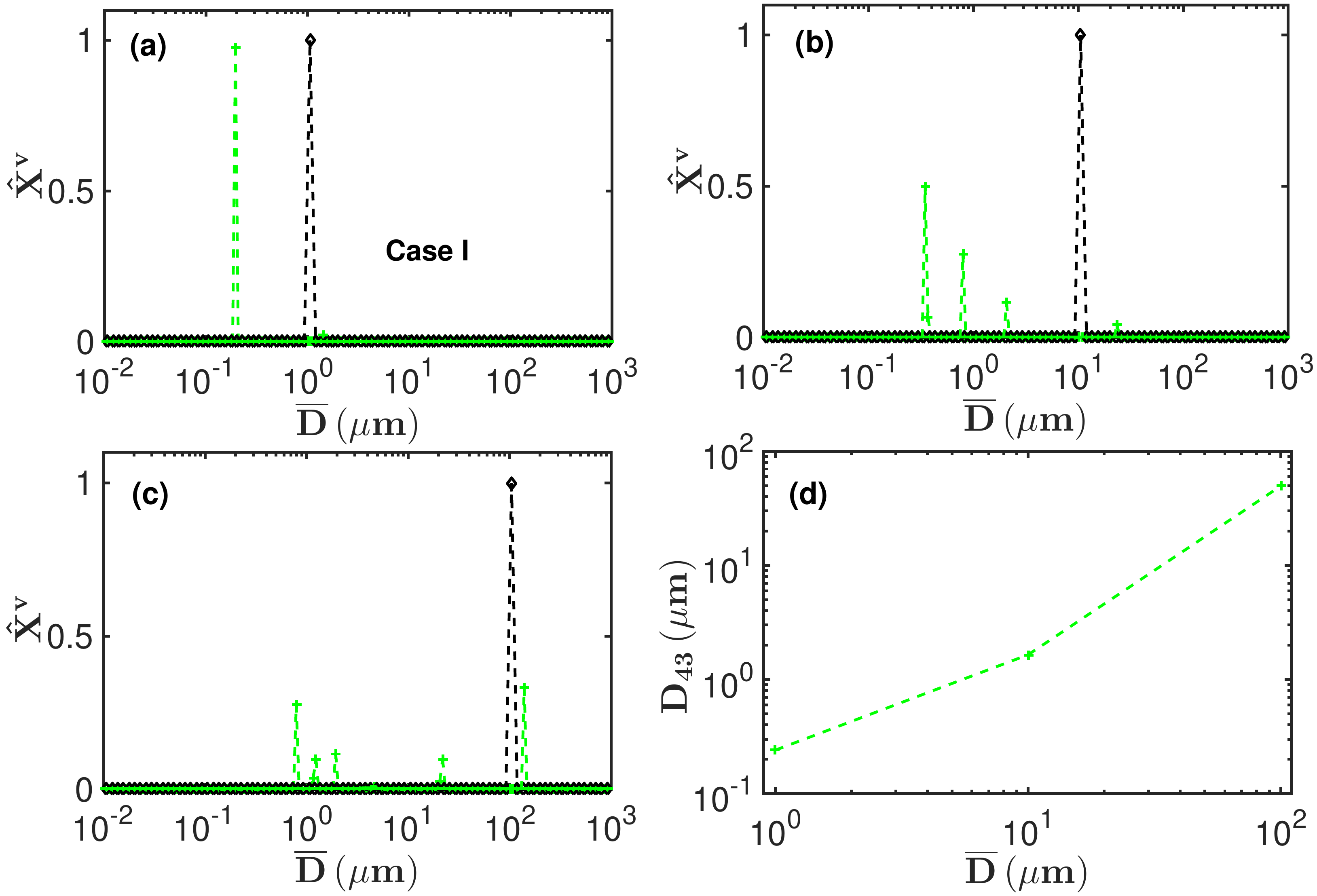}}
\caption{
Comparison of simulated and estimated size distributions. Black diamonds: monodisperse  distributions of circular cross-sectional diameters for needle-like particles of (a)\,$\overline{D}=1\mu$m, (b)\,$\overline{D}=10\mu$m and (c)\,$\overline{D}=100\mu$m for Case I. Green crosses: multimodal volume based distributions of diameters of spherical particles which give the best fit to the scattering intensities (shown in Fig. \ref{fig3}) of the needle-like particles whose distribution of cross-sectional diameters are shown by the black diamonds in (a), (b) and (c). The best fits are obtained by solving the optimisation problem in Eq. \eqref{eq6} with the spherical model. (d)\,$D_{43}$ values estimated from the volume based PSDs obtained with the spherical model. The needle-like particles were modelled as infinitely long cylinders
}
\label{fig4}
 \end{figure}

 \begin{figure}[tbh]
\centerline{\includegraphics[width=0.5\textwidth]{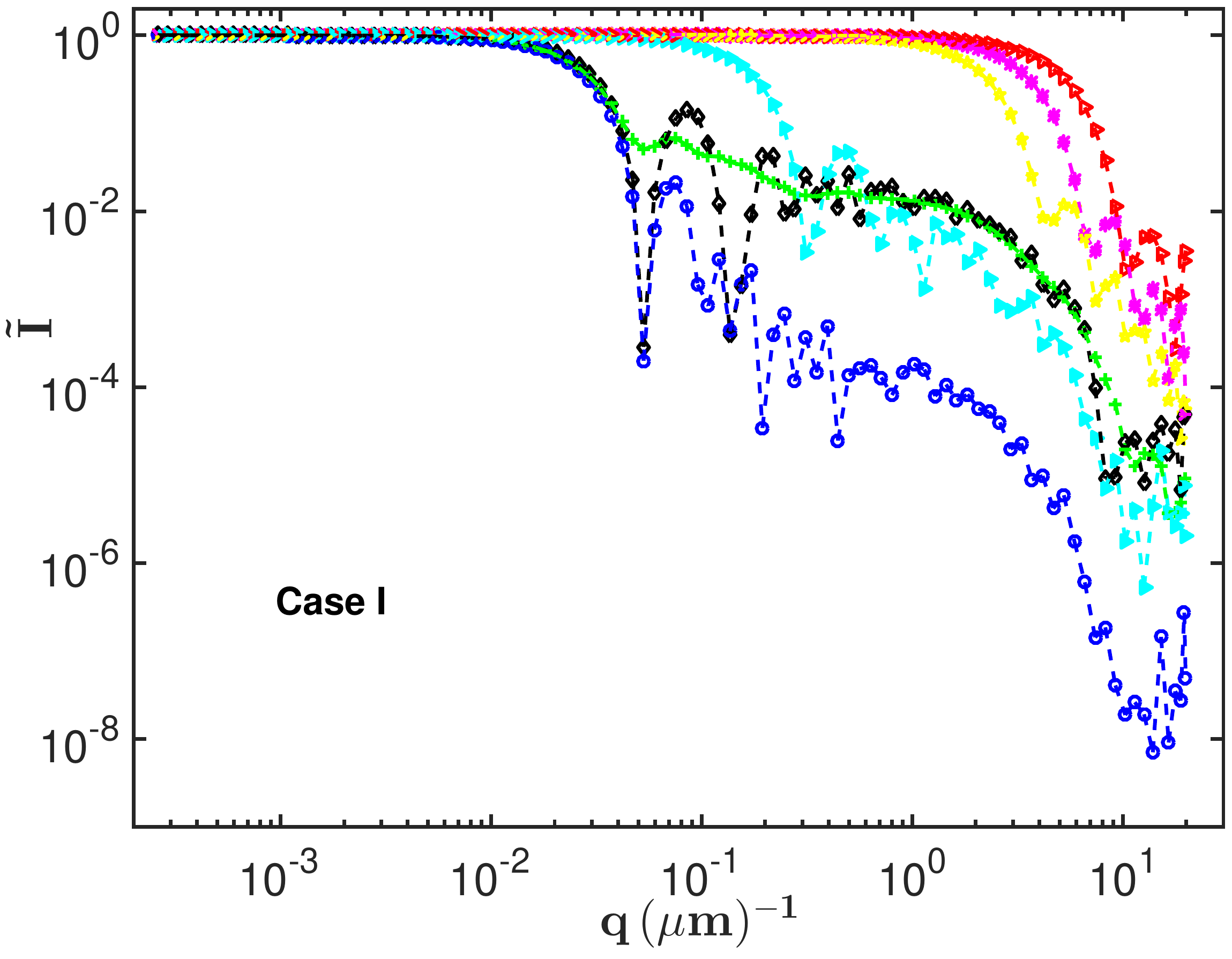}}
\caption{
Calculated scattering intensities for monodisperse populations of spherical particles of diameters
 $\overline{D}=0.8\mu$m (red triangles), $\overline{D}=1.2\mu$m (magenta filled symbols), $\overline{D}=1.9\mu$m (yellow pentagrams), $\overline{D}=21\mu$m (cyan filled triangles) and $\overline{D}=140\mu$m (blue circles).
 The green crosses are the superposition of the corresponding scattering intensities
 (weighted by the number fractions estimated from the volume fractions shown by the green crosses in Fig. \ref{fig4}(c)). 
 The black diamonds are the scattering intensity for needle-like particles of circular cross-sectional diameter $\overline{D}=100\mu$m. 
The needle-like particles were modelled as infinitely long cylinders
}
\label{fig5}
 \end{figure}

The calculated scattering intensities for monodisperse needle-like particles with the diameter of circular cross-section of $\overline{D}=1\mu$m, $10\mu$m,  and 100$\mu$m are shown in Fig. \ref{fig3} (black diamonds). The blue circles in Fig. \ref{fig3} show the calculated scattering intensity for monodisperse spherical particles with diameter equal to the diameter of circular cross-section of needle-like particles. It can be seen in Fig. \ref{fig3}(a) that for $\overline{D}=1\mu$m both scattering patterns are quite similar for lower $q$ values, while in the high $q$ ($q\gtrsim 5\mu$m$^{-1}$) region the scattering intensity for spheres is much lower than that for the needle-like particles. However, we can see that the scattering intensity pattern for monodisperse needle-like particles can be reasonably well fitted using scattering from a multimodal distribution of spherical particles. The fitted curve shown by the green crosses in Fig. \ref{fig3}(a) is obtained by solving the optimisation problem in Eq. \eqref{eq6}. 

In order to quantify the degree of fit, we compute the $L_2$ norm ($L_2^{fit}$) of the difference between the scattering intensities of the needle-like particles (modelled as infinitely long cylinders) and that of the fitted curve from the spherical model. This gives $L_2^{fit} = 0.142$. This value is nearly one third of the $L_2$ norm ($L_2^{sphere}$) of the difference between the scattering intensities of the needle-like particles and that of the spherical particles whose diameters are equal to those of the circular cross-sectional diameter of the infinitely long cylinder. The value of $L_2^{sphere}$ is 0.415 for the case of $\overline{D}=1\mu$m.

The multimodal PSD (in terms of volume fractions $\hat{X}^v$) of spherical particles is shown in Fig. \ref{fig4}(a). In order to achieve a reasonable fit 
(quantified by the $L_2^{fit}$ value), 
the estimated PSD needs to contain at least two modes, with 98\% (by volume) of particles with diameter $\approx 0.2\mu$m and 2\% (by volume) of particles with diameter $\approx 1.4\mu$m. This can be understood by considering the difference in shapes between scattering patterns of spheres and thin cylinders.

In the low $q$ region, the normalised scattering intensity $\tilde{I}$ is independent of $q$ regardless of particle shape. In the Guinier region, the scattering intensity starts decreasing with increasing $q$ and can be approximated as $\tilde{I}(q)/\tilde{I}(0)=1-q^2R^2_g/3$ \cite{Sorensen2000}, where $R_g$ is the radius of gyration of the particles and $\tilde{I}(0)$ is the zero $q$ limit of $\tilde{I}$ which is indicated as $I_1$ in Eq. \eqref{eq4}. The Guinier region extends towards $qR_g\approx 1$ where a noticeable downturn of the scattering intensity can be observed. For spherical particles $R_g=\overline{D}\sqrt{(3/20)}$ so that the downturn in the scattering intensity can be seen when $q\approx (2/\overline{D})\sqrt{(5/3)}$ (i.e., at $q\approx 3\mu$m$^{-1}$ for $\overline{D}=1\mu$m). At large $q$ values the scattering intensity transits into the Porod region where it decays with an exponent depending on the nature of the particle surface 
(for smooth surfaces $\tilde{I}\sim q^{-4}$ \cite{Sorensen2001}).

It can be seen in Fig. \ref{fig3}(a) that the scattering intensity pattern for the monodisperse needle-like particles
(modelled as infinitely long cylinders)
 of circular cross-sectional diameter $\overline{D}=1\mu$m (black diamonds) turns down at slightly lower $q$ values than for spheres of the same diameter. Therefore a slightly larger sphere diameter is needed to reproduce the Guinier region of a thin cylinder and the estimated PSD (green crosses in Fig. \ref{fig4}(a)) contains a peak at a particle diameter slightly larger than $1\mu$m. The scattering intensity pattern for thin cylinders of circular cross-sectional diameter $\overline{D}=1\mu$m initially decays faster than that of spheres of the same diameter at higher $q$ values (Fig. \ref{fig3}(a)). However, the decrease of the scattering intensity pattern for the thin cylinders slows down considerably as $q$ increases, whereas the corresponding scattering intensity for spheres continue to decrease with increasing $q$. Hence, the scattering pattern of a thin cylinder cannot be fitted with just that of a single sphere. In order to get a reasonable fit 
 (judging by the $L_2^{fit}$ value),
  it is necessary to include additional scattering at larger $q$ values, which can only come from much smaller spherical particles and the resulting PSD will become multimodal. This can be seen as a second peak at $\overline{D}\approx 0.2\mu$m in the estimated PSD in Fig. \ref{fig4}(a). This peak is a pure artefact in the estimated PSD as it does not correspond to any physical length scale related to the particle size and arises solely from the difference of scattering patterns between cylinders and spheres.
The same phenomenon can be seen for other cylinder sizes as shown in Figs. \ref{fig3}(b) and \ref{fig4}(b) (for circular cross-section diameter of $\overline{D} = 10\mu$m and $L_2^{fit}=0.828, L_2^{sphere}=1.318$) and Figs. \ref{fig3}(c) and \ref{fig4}(c) (for circular cross-section diameter of $\overline{D} = 100\mu$m and and $L_2^{fit}=0.181, L_2^{sphere}=0.521$).
The consequence of these artificial particle size modes are the grossly under-estimated $D_{43}$ values for the respective particle sizes as seen in Fig. \ref{fig4}(d).

 \begin{figure}[tbh]
\centerline{\includegraphics[width=0.8\textwidth]{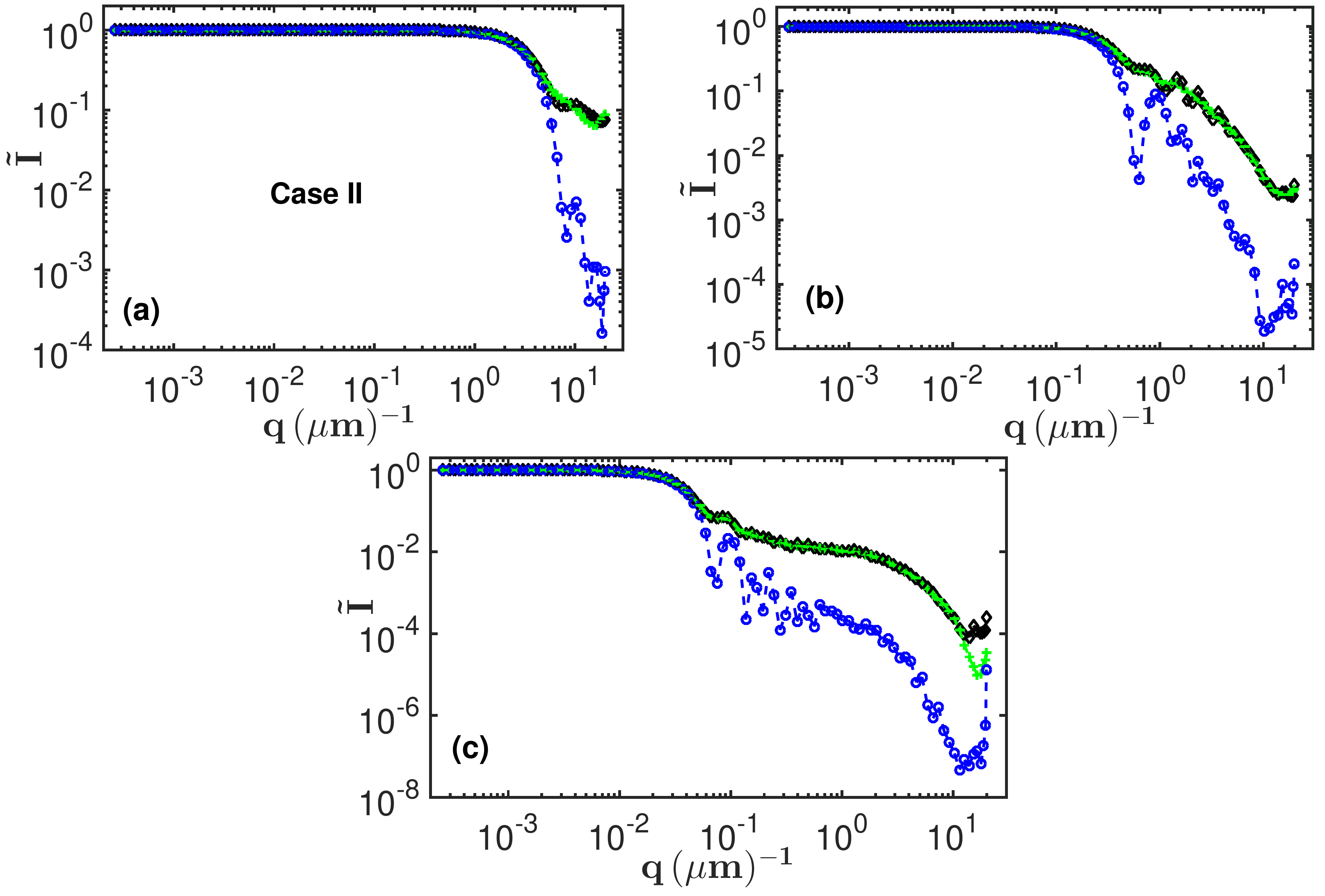}}
\caption{
Comparison of normalised scattering intensity patterns for needle-like and spherical particles of various sizes. Similar to Fig. \ref{fig3} but for Case II.
}
\label{fig6}
 \end{figure}

 \begin{figure}[tbh]
\centerline{\includegraphics[width=0.8\textwidth]{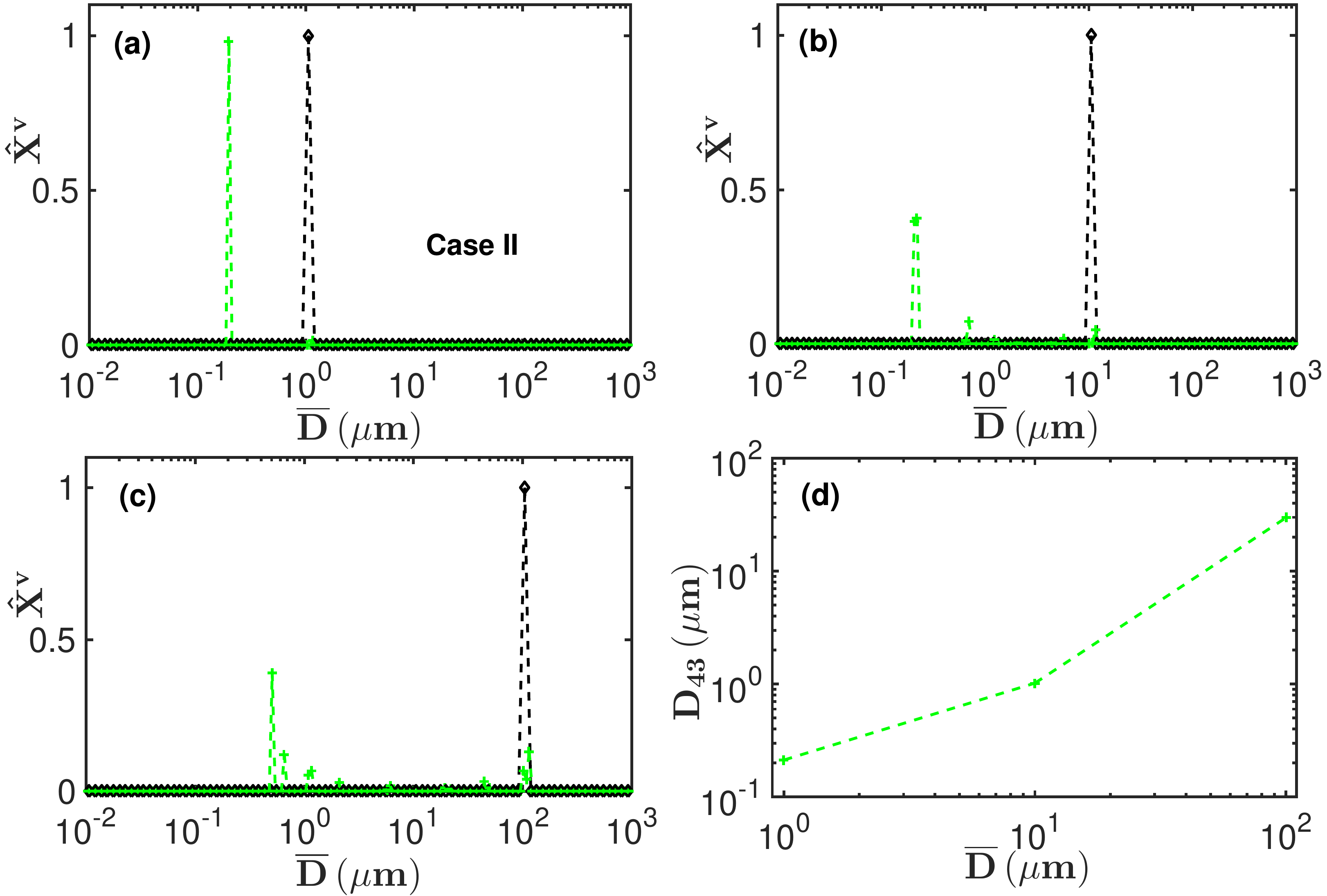}}
\caption{
Multimodal distributions of diameter of spherical particles corresponding to best fits of scattering patterns of monodisperse needle-like particles. Similar to Fig. \ref{fig4} but for Case II.
}
\label{fig7}
 \end{figure}

To further understand
that using the spherical model requires
 a multimodal PSD to fit the scattering intensity of the needle-like particles, consider the case of the scattering intensity of needle-like particles 
 (modelled as infinitely long cylinders)
 with circular cross-sectional diameter $\overline{D} = 100\mu$m shown by the black diamonds in Fig. \ref{fig5}. The PSD estimated by fitting (solving the 
optimisation
 problem in Eq. \eqref{eq6}) the spherical model to this scattering intensity contains major peaks at $\overline{D} = 0.8\mu$m, $\overline{D} = 1.2\mu$m, $\overline{D} = 1.9\mu$m, $\overline{D} = 21\mu$m and $\overline{D} = 140\mu$m as shown by the green crosses in Fig. \ref{fig4}(c). A scattering intensity corresponding to the fitted curve shown by the green crosses in Fig. \ref{fig3}(c) can be constructed by performing a superposition of the scattering intensities for monodisperse spherical particles of sizes $\overline{D} = 0.8\mu$m, $\overline{D} = 1.2\mu$m, $\overline{D} = 1.9\mu$m, $\overline{D} = 21\mu$m and $\overline{D} = 140\mu$m shown in Fig. \ref{fig5}.

 The superposition is performed by weighting the scattering intensities by their respective number fractions estimated as $\hat{X}_s = \hat{X}_s^v/\overline{D}^3_s$ where subscript $s$ represents particle size and $\hat{X}_s^v$ is the volume fraction corresponding to each size as shown in Fig. \ref{fig4}(c). The normalised scattering intensity obtained by this superposition procedure is shown by the green crosses in Fig. \ref{fig5}, and it has a good fit ($L_2^{fit}=0.182$) to the original scattering intensity (black open diamonds in Fig. \ref{fig5}) for the needle-like particles (modelled as infinitely long cylinders) of circular cross-sectional diameter $\overline{D} = 100\mu$m. This fit cannot be obtained if the weighted scattering intensities from the smaller particle sizes of
 $\overline{D} = 0.8\mu$m, $\overline{D} = 1.2\mu$m, $\overline{D} = 1.9\mu$m and $\overline{D} = 21\mu$m are not included in the superposition.

\subsection{Case II: random rotations}
\label{ssec4-2}

The Case II where the needle-like particles are able to make random rotations is analysed in this section. As in Case I, the scattering intensity pattern of the needle-like particles will be simulated with the model for infinitely long cylinders, while the inverse problem will be solved with the model for spherical particles. In this case, the scattering
intensities are computed for various values of incident angle from $1^{\circ}$ to $90^{\circ}$. Then the overall scattering intensity is obtained by averaging these scattering intensities over all angles.

Compared to Case I, the Guinier region of the scattering intensity (blue circles in Fig. \ref{fig6}) for spherical particles of diameter $\overline{D}=1\mu$m matches that of needle-like particles (black diamonds in Fig. \ref{fig6}) of the same circular cross-sectional diameter much more closely. This leads the spherical model to estimate the largest particle size with the diameter which is very close to the circular cross-sectional diameter of the needle-like particles (modelled as infinitely long cylinders) as shown by the green crosses in Fig. \ref{fig7}. However, similar to Case I, the decay of the scattering intensity pattern for needle-like particles slows down at higher $q$ values when compared with those of spheres of the same cross-sectional diameter as seen in Fig. \ref{fig6}. Hence as in Case I, in order to get a reasonable fit (based on the $L_2^{fit}$ value), it is necessary to include additional scattering at larger $q$ values, which must come from particles with much smaller diameters and the resulting PSD thus become multimodal, although the smaller size peaks are artefacts in the estimated PSD. As before, the consequence of these additional peaks introduced into the estimated PSD by the spherical model is that the estimated $D_{43}$ values can be significantly lower that the actual cross-sectional diameter of needle-like particles as seen in Fig. \ref{fig7}(d).

\section{Conclusions}
\label{sec5}

We have demonstrated the origin of artefacts which arise due to applying a model for scattering by spherical particles to solve the inverse problem for laser diffraction when the scattering intensity pattern comes from a population of needle-like particles. The scattering intensity patterns for the needle-like particles were simulated with the model for infinitely long cylinders using a monodisperse distribution (in terms of the circular cross-sectional diameters) of these cylinders. Our results show that using the scattering model for spherical particles it is possible to find a good fit for the scattering intensity patterns of needle-like particles, but the resulting estimated PSD is not necessarily representative of actual cross-sectional diameters of the needle-like particles. The estimated PSDs are typically multimodal with the largest size mode close to the actual cross-sectional diameter but additional smaller size modes are not physical. These modes are mathematical artefacts arising from different shapes of scattering patterns of spheres and thin cylinders. Needle-like particles of various lengths would be expected to  give rise to similar effects which would depend on the aspect ratio of the particles. It can be expected that the same issue applies for particles of other shapes, such as platelets. 

 This situation is unavoidable as long as the scattering model for spherical particles is used to fit the data from particles of strongly non-spherical shapes as it is possible to fit essentially any scattering pattern with the scattering models for a polydiperse population of spheres. This approach is used in commercial laser diffraction instruments, and it can lead to misleading conclusions about the PSD of the particles under analysis, in terms of unrealistic multimodal distributions and underestimating the mean particle size in terms of volume weighted mean diameter values. Therefore, the way forward is to obtain information about particle shape and apply appropriate scattering models for particles to be analysed.

\section*{Acknowledgement}

This work was performed within the UK EPSRC funded project \\
 (EP/K014250/1) `Intelligent Decision Support and Control Technologies for Continuous Manufacturing and Crystallisation of Pharmaceuticals and Fine Chemicals' (ICT-CMAC). The authors would like to acknowledge financial support from EPSRC, AstraZeneca and GSK. The authors are also grateful for useful discussions with industrial partners from AstraZeneca, GSK, Mettler-Toledo, Perceptive Engineering and Process Systems Enterprise. The authors also wish to thank Thomas McGlone and Vaclav Svoboda for providing the images in Fig.\ref{fig1}.

 \newpage
 
 \setcounter{section}{0}

 \setcounter{equation}{0}
 
 \setcounter{figure}{0}

 \setcounter{footnote}{0}
 
 \begin{center}
 \LARGE{\textbf{Supplementary Information}}
 \end{center}
  
\section{Calculating scattering intensity}
\label{sup_sec1}

As discussed in section 2 of the main text, the intensity $\mathbf{I}$ of scattered light is related to that of the incident light $\mathbf{I}_0$ by means of the phase matrix. The calculation is performed by considering the detector system (sketched in Fig. 2(a) of the main text) in which a monochromatic light  with wave vector $\mathbf{k}_i$ is incident on a particle of arbitrary size and shape. The scattered light with wave vector $\mathbf{k}_s$ is then collected at different angles $\theta$ to the direction of propagation of the incident light by an array of detectors as depicted in Fig. 2(a) of the main text. Both the incident and scattered light have components parallel and perpendicular to the scattering plane (the plane containing the incident and scattered light) \cite{Bohren1983}. The scattering wave vector $\mathbf{q}$ is the difference between the incident and scattered wave vectors as sketched in Fig. 2(a) of the main text. 

The general information concerning the intensity and state of polarisation of both the incident and scattered light is contained in the Stokes parameters $I, Q, U, V$ \cite{Bohren1983,Mishchenko2000}, where $I$ is the light flux or intensity, $Q$ and $U$ describe the state of linear polarisation and $V$ describes the state of circular polarisation. The Stokes parameters of the incident light are related to those of the scattered light by means of the scattering matrix or phase matrix $\mathbf{Z}$ as \cite{Bohren1983,Mishchenko2000}
\begin{equation}
 \left(\begin{array}{c}
I_s \\
Q_s \\
U_s \\
V_s
\end{array} \right) = \frac{1}{k^2r^2} \left(\begin{array}{cccc}
Z_{11} & Z_{12} & Z_{13} & Z_{14} \\
Z_{21} & Z_{22} & Z_{23} & Z_{24} \\
Z_{31} & Z_{32} & Z_{33} & Z_{34} \\
Z_{41} & Z_{42} & Z_{43} & Z_{44} 
\end{array} \right) \left( \begin{array}{c}
I_i \\
Q_i \\
U_i \\
V_i
\end{array} \right),
\label{eqs1}
\end{equation}
where $k=2\pi/\lambda$ is the wave number and $r$ is distance in the radial direction. The Stokes parameters of the scattered light are indicated by the subscript $s$, while those of the incident wave are indicated by the subscript $i$ in Eq. \eqref{eqs1}. The elements $Z_{ij}, ij=1,2,3,4$ of the phase matrix $\mathbf{Z}$ are related to the elements of the amplitude matrix $\mathbf{S}$. The amplitude matrix relates the incident electric field $\mathbf{E}_i$ to the scattered electric field $\mathbf{E}_s$ as \cite{Bohren1983,Mishchenko2000}
\begin{equation}
 \left(\begin{array}{c}
E_{\parallel s} \\
E_{\perp _s} 
\end{array} \right) = \frac{e^{ik(r-z)}}{-ikr} \left(\begin{array}{cc}
S_2 & S_3 \\
S_4 & S_1 
\end{array} \right) \left( \begin{array}{c}
E_{\parallel _i} \\
E_{\perp _i} \end{array}\right),
\label{eqs2}
\end{equation}
where the incident light is taken to propagate in the $z$ direction. The elements $S_i, i=1,2,3,4$ of the amplitude matrix are functions of the scattering angle $\theta$, the size of the particle $D$ (the size of a particle is taken to be the diameter of spherical particles or the diameter of circular cross section of needle-like particles which are modelled as infinitely long cylinders in this work) and the refractive index $N_R$ of the particle.

Most commercial laser diffraction instruments only measure the intensity of scattered light from an incident unpolarised light. Hence it is assumed in this work that there is no polarisation of the scattered light so that the Stokes parameters $Q=U=V=0$.  Thus Eq. \eqref{eqs1} reduces to
\begin{equation}
I_s = \frac{Z_{11}}{k^2r^2}I_i,
\label{eqs3}
\end{equation}
where $Z_{11}$ is given by \cite{Bohren1983}
\begin{equation}
Z_{11} = \frac{1}{2}\left(|S_1|^2 + |S_2|^2 + |S_3|^2 + |S_4|^2\right).
\label{eqs4}
\end{equation}

In a population of particles of different sizes and shapes, the phase matrix for the population will be the sum of the phase matrices of the individual particles in the population. Then the phase matrix for the population is given as \cite{Bohren1983,Mishchenko2000}
\begin{equation}
\mathbf{Z} = \sum_{k=1}^{\mathcal{N}}{\mathbf{Z}_k},
\label{eqs5}
\end{equation}
where $\mathcal{N}$ is the number of particles in the population. The sum in Eq. \eqref{eqs5} holds provided that the particles are randomly positioned and oriented, and that the particles are sufficiently spaced such that there is no multiple scattering. Hence the intensity of scattered light from the population of particles will be given as
\begin{equation}
I = \frac{I_0}{k^2r^2}\sum_{k=1}^{\mathcal{N}}Z_{11,k},
\label{eqs6}
\end{equation}
where $I_0$ and $I$ are the intensities of the incident and scattered light respectively and $Z_{11,k}$ is the contribution to the $Z_{11}$ component of the phase matrix from particle $k$.
The conditions required for Eq. \eqref{eqs5} to hold can be met in commercial laser diffraction instruments, hence the conditions will be assumed to apply in this work.

Here, the polydispersed population of particles shall be assumed to consist of particles of the same shape and refractive index but different sizes. The distribution of particle sizes is characterised by a probability density function $n(D)$ such that $n(D)dD$ is the probability of finding particles with diameters between $D$ and $D+dD$. If the diameters of the particles are discretised and grouped into $N$ geometrically spaced size classes such that the characteristic diameter of particles whose diameters lie between $D_i$ and $D_{i+1}$ ($i=1,2,\ldots,N$) is given by $\overline{D}_i = \sqrt{D_iD_{i+1}}$, then a discretised particle size distribution (PSD) $X(\overline{D}_i)$ can be defined such that $X(\overline{D}_i)$ is the number of particles whose diameters lie between $D_i$ and $D_{i+1}$.

Then the number of particles $\mathscr{N}_i$ with characteristic diameter $\overline{D}_i$ is given by $\mathscr{N}_i = X(\overline{D}_i)$. Then Eq. \eqref{eqs6} can be rewritten as
\begin{equation}
I(\theta_j) = \frac{I_0}{k^2r^2}\sum_{i=1}^{N}\mathscr{N}_iZ_{11}(\theta_j,\overline{D}_i)  = \frac{I_0}{k^2r^2}\sum_{i=1}^{N}Z_{11}(\theta_j,\overline{D}_i)X(\overline{D}_i),
\label{eqs7}
\end{equation}
where the scattering angle $\theta$ has been discretised into $j=1,2\ldots,M$ angular positions and $Z_{11}(\theta_j,\overline{D}_i)$ is the averaged phase matrix component for particles whose sizes lie between $D_i$ and $D_{i+1}$.
Since the magnitude of the scattering wave vector $q$ is a function of angle (as defined in Eq. (1) of the main text), then the scattering intensity and component $Z_{11}$ of the phase matrix in Eq. \eqref{eqs7} can be rewritten as functions of $q$ (discretised) as 
\begin{equation}
I(q_j) = \frac{I_0}{k^2r^2}\sum_{i=1}^{N}Z_{11}(q_j,\overline{D}_i)X(\overline{D}_i).
\label{eqs8}
\end{equation}
Writing $I(q_j)$ as $I_j$, $Z_{11}(q_j,\overline{D}_i)$ as $\tilde{Z}_{j,i}$ and $X(\overline{D}_i)$ as $X_i$, then Eq. \eqref{eqs8} can be written as a matrix equation as
\begin{equation}
I_j = \mathbf{I} = \frac{I_0}{k^2r^2}\sum_{i=1}^{N}\tilde{Z}_{j,i}X_i = \frac{I_0}{k^2r^2}\mathbf{\tilde{Z}X}.
\label{eqs9}
\end{equation}

The scaling factor in Eq. \eqref{eqs9} is removed by rescaling the scattering intensity by the scattering intensity $I_1$ to give the rescaled scattering intensity $\mathbf{\tilde{I}}_j$ as
\begin{equation}
\tilde{I}_j = \frac{I_j}{I_1} = \frac{\sum_{i=1}^{N}\tilde{Z}_{j,i}X_i}{\sum_{i=1}^{N}\tilde{Z}_{1,i}X_i}. 
\label{eqs10}
\end{equation}

\subsection{Population of spherical particles}
\label{sup_ssec2-1}

The components $S_3 = S_4 = 0$ in the amplitude matrix defined in Eq. \eqref{eqs2} in the case of spherical particles \cite{Bohren1983}. The components $S_1$ and $S_2$ are defined as \cite{Bohren1983}
\begin{subequations}
\begin{align}
S_1 = & \sum_n{\frac{2n+1}{n(n+1)}(a_n\pi_n + b_n\tau_n)}\\
S_2 = & \sum_n{\frac{2n+1}{n(n+1)}(a_n\tau_n + b_n\pi_n)},
\end{align}
\label{eqs11}
\end{subequations}
where the quantities $\pi_n$ and $\tau_n$ are functions of the scattering angle $\theta$. They can be obtained by the following recurrence formulas \cite{Bohren1983}
\begin{subequations}
\begin{align}
\pi_n = & \frac{2n-1}{n-1}\mu\pi_{n-1} - \frac{n}{n-1}\pi_{n-2}\\
\tau_n = & n\mu\pi_n - (n+1)\pi_{n-1},
\end{align}
\label{eqs12}
\end{subequations}
where $\mu = \cos(\theta)$, $\pi_0 = 0$ and $\pi_1=1$. The series in Eqs. \eqref{eqs11} and \eqref{eqs12} are truncated after $n_c$ terms; the value of which is related to the size of the particle \cite{Bohren1983}. The scattering coefficients $a_n$ and $b_n$ are functions of the particle size and refractive index. They are obtained by solving the Maxwell equations for a spherical particle with appropriate boundary conditions\footnote{See \cite{Bohren1983} for details of the calculations of the scattering coefficients. The numerical calculations of these scattering coefficients (for spherical particles) has been implemented in various programming languages. The Matlab implementation by \cite{Maetzler2002} is used in this work.}. Using Eq. \eqref{eqs4}, then the $Z_{11}$ component of the phase matrix can be calculated and subsequently the rescaled scattering intensity of the population can be calculated using Eq. \eqref{eqs10}.

\subsection{Population of needle-like particles}
\label{sup_ssec2-2}

 \begin{figure}[tbh]
\centerline{\includegraphics[width=0.5\textwidth]{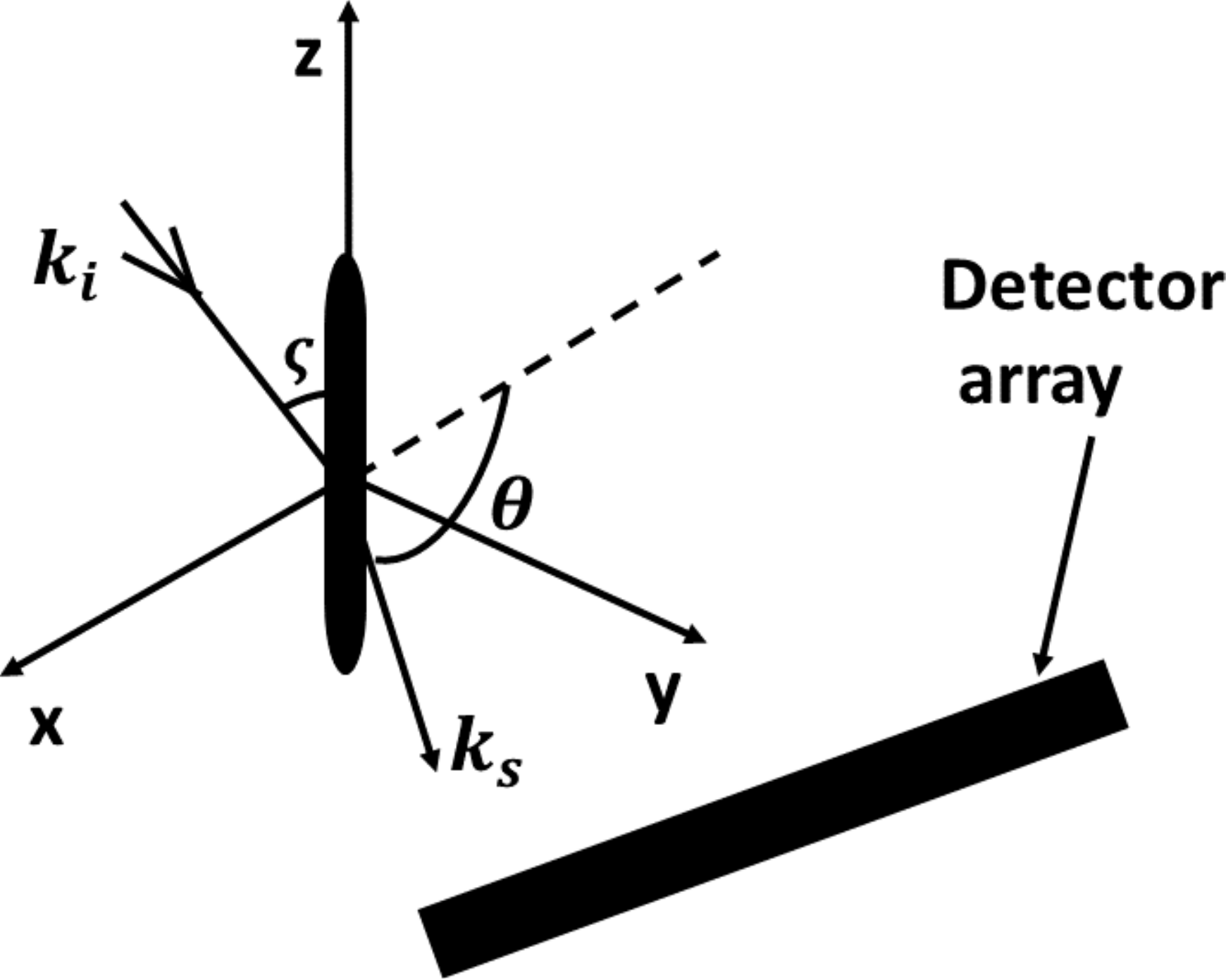}}
\caption{A schematic for light scattering by an infinitely long cylinder.}
\label{figs1}
 \end{figure}

The scattering intensity of needle-like particles is approximated with that for infinitely long cylinders in this work. The sketch of such an infinitely long cylinder is shown in Fig. \ref{figs1}. The axis of the cylinder lies along the $z$-axis. The incident light (with wave vector $\mathbf{k}_i$) which is contained in the $x-z$ plane makes an angle $\zeta$ with the cylinder axis, while the scattered light (with wave vector $\mathbf{k}_s$) is contained in the $x-y$ plane and makes an angle $\theta$ with the negative $x$-axis. The scattered light is measured in the $x-y$ plane by an array of detectors as shown in Fig. \ref{figs1}. The scattering plane contains the cylinder axis and scattered light \cite{Bohren1983}. 

The amplitude matrix relates the incident electric field $\mathbf{E}_i$ to the scattered electric field $\mathbf{E}_s$ as \cite{Bohren1983}
\begin{equation}
 \left(\begin{array}{c}
E_{\parallel s} \\
E_{\perp _s} 
\end{array} \right) = e^{i3\pi/4}\sqrt{\frac{2}{\pi kr\sin(\zeta)}}e^{ik(r\sin(\zeta) - z\cos(\zeta))}
 \left(\begin{array}{cc}
T_1 & -T_3 \\
T_3 & T_2 
\end{array} \right) \left( \begin{array}{c}
E_{\parallel _i} \\
E_{\perp _i} \end{array}\right),
\label{eqs13}
\end{equation}
The quantities $T_i,i=1,2,3$ in Eq. \eqref{eqs13} correspond to the quantities $S_i,i=1,2,3,4$ in Eq. \eqref{eqs2} for the general case. The quantities $T_i,i=1,2,3$ are given as \cite{Bohren1983}
\begin{subequations}
\begin{align}
T_1 = & \hat{b}_{0I} + 2\sum_{n=1}^{\infty}{\hat{b}_{nI}\cos{(n\theta)}} \\
T_2 = & \hat{a}_{0II} + 2\sum_{n=1}^{\infty}\hat{a}_{nII}\cos{(n\theta)} \\
T_3 = & -2i\sum_{n=1}^{\infty}{\hat{a}_{nI}\sin{(n\theta)}}. 
\end{align}
\label{eqs14}
\end{subequations}
The scattering coefficients $\hat{b}_{0I}, \hat{b}_{nI}, \hat{a}_{0II}, \hat{a}_{nII}, \hat{a}_{nI}$ (similar to the case of $a_n$ and $b_n$ in Eq. \eqref{eqs11}) are functions of particle size (the diameter of the infinitely long cylinder) and refractive index. They are obtained by solving the Maxwell equations for the cylindrical geometry with appropriate boundary conditions\footnote{See \cite{Bohren1983} for a detailed discussion of the calculation of these coefficients. The numerical computations of these coefficients were carried out in Matlab in this work.}.

Using the components $T_i,i=1,2,3$ of the amplitude matrix for the infinitely long cylinder, then the $Z_{11}$ component of the phase matrix can be calculated as 
\begin{equation}
Z_{11} = \frac{1}{2}\left(|T_1|^2 + |T_2|^2 + 2|T_3|^2\right).
\label{eqs15}
\end{equation}
Subsequently the rescaled scattering intensity in Eq. \eqref{eqs10} can be calculated for the infinitely long cylinders that are used to represent needle-like particles in this work.

\section{Number and volume based PSD}
\label{sup_sec2}

The PSD defined in Eq. \eqref{eqs7} which is calculated by solving the least square problem in Eq. (6) of the main text is number based. The number based PSD $X_i$ is defined as an exponential function of the parameter $\gamma_i$ as
\begin{equation}
X_i = e^{\gamma_i}, i=1,2,\ldots,N.
\label{eqs16}
\end{equation}
Then the optimisation problem given in Eq. (6) of the main text is solved by searching for $\gamma_i$ (using the Levenberg-Marquardt algorithm as implemented in Matlab) which minimises the objective function $f$ given in Eq. (6) of the main text. As the Levenberg-Marquardt method is gradient based and the objective function in Eq. (6) of the main text contains local minima, then a multi-start strategy \cite{Aster2013} is used to search for a  solution which is close to the global minimum. This involves using different random starting solutions for $\gamma_i$, then the solution for which the $L_2$ norm given as 
\begin{equation}
\|\mathbf{\tilde{I}}^{\ast} - \mathbf{\tilde{I}}\|_2 = \sqrt{\sum_{j=1}^M{\left[\tilde{I}_j^{\ast} -  \tilde{I}_j\right]^2}}
\label{eqs17} 
\end{equation}
is minimum is then chosen as the optimum (the scattering intensity $\mathbf{\tilde{I}}^{\ast}$ is that of the needle-like particles which are modelled as infinitely long cylinders, while the scattering intensity $\mathbf{\tilde{I}}$ is that of spherical particles). Subsequently the number based PSD $X_i$ in Eq. \eqref{eqs16} is calculated from this optimum solution for $\gamma_i$. 

However, as commercial laser diffraction instruments typically report a volume based PSD, then it is necessary to calculate a corresponding volume based PSD. This can be achieved as follows. Consider the scattering intensity $\overline{I}_j$ (obtained in a manner similar to the case of $I_j$ in Eq. \eqref{eqs9}) given as
\begin{equation}
\overline{I}_j = \alpha\sum_{i=1}^{N}{\tilde{Z}_{ji}\hat{X}_i},
\label{eqs18}
\end{equation}
where 
\begin{equation}
\alpha = \frac{I_0}{k^2r^2}
\label{eqs19}
\end{equation}
and
\begin{equation}
\hat{X}_i = \frac{X_i}{\sum_{i=1}^{N}{X_i}}.
\label{eqs20}
\end{equation}

The scattering intensity $\overline{I}_j$ is associated with the number based PSD $X_i$ by means of Eq. \eqref{eqs18}. However, the scattering intensity $\overline{I}_j$ can also be associated with the volume based PSD $X^v_i$ (the total volume of particles with diameters between $D_i$ and $D_{i+1}$) by writing
\begin{equation}
\overline{I}_j = \alpha\sum_{i=1}^{N}{\overline{Z}_{ji}\overline{X}_i^v},
\label{eqs21}
\end{equation}
where
\begin{equation}
\overline{Z}_{ji} = \frac{\tilde{Z}_{ji}}{D_i^3},
\label{eqs22}
\end{equation}

\begin{equation}
\overline{X}_i^v = \hat{X}^v_i\sum_{i=1}^{N}{\hat{X}_iD^3_i}
\label{eqs23}
\end{equation}
and
\begin{equation}
\hat{X}_i^v = \frac{\hat{X}_iD^3_i}{\sum_{i=1}^{N}{\hat{X}_iD^3_i}}.
\label{eqs24}
\end{equation}

Then the scattering intensity $\overline{I}$ can be normalised to remove the scaling factor $\alpha$ (as in the case of Eq. \eqref{eqs10}) as
\begin{equation}
\hat{I}_j = \frac{\sum_{i=1}^{N}{\overline{Z}_{ji}\overline{X}_i^v}}{\sum_{i=1}^{N}{\overline{Z}_{1i}\overline{X}_i^v}}.
\label{eqs25}
\end{equation}
Hence Eq. \eqref{eqs25} becomes the forward problem for the volume based PSD\footnote{The association of the scattering intensity $\overline{I}_j$ to the volume based PSD given in Eq. \eqref{eqs21} can only be made for particles with defined volume, for example spherical particles. However, as the sizes of the infinitely long cylinders used in this work are monodispersed, then the number based PSDs will coincide with the corresponding volume based PSDs of the infinitely long cylinders even though the volume of an infinitely long cylinder is not defined.} similar to the case of Eq. \eqref{eqs11} for the number based PSD. 

The volume based PSD $X_i^v$ can then be calculated by solving a weighted least square problem similar to the case given in Eq. (6) of the main text. A similar approach has previously been implemented for the focused beam reflectance measurement sensor data \cite{Agimelen2015,Agimelen2016}. 

The method of computing the volume based PSD $X^v_i$ is carried out as follows. Obtain the number based PSD $X_i$ whose corresponding scattering intensity $\tilde{I}_j$ gives the best fit to the scattering intensity $\tilde{I}^{\ast}_j$ of the needle-like particles as judged by the $L_2$ norm in Eq. \eqref{eqs17} . Then using this optimum number based PSD, construct the scattering intensity $\overline{I}_j$ given in Eq. \eqref{eqs18}. Then normalise the scattering intensity $\overline{I}_j$ to obtain the scattering intensity $\hat{I}_j$ in Eq. \eqref{eqs25} as $\hat{I}_j = \overline{I}_j/\overline{I}_1$, where $\overline{I}_1$ is the zero $q$ limit of $\overline{I}_j$. This scattering intensity $\hat{I}_j$ is associated with the volume based PSD $X^v_i$ defined in Eq. \eqref{eqs25}. When $\hat{I}_j$ is computed for needle-like particles, then it is written as $\hat{I}_j^{\ast}$.

To obtain the volume based PSD, search for values of the parameters $\gamma^v_i$ for which the volume based PSD $\overline{X}^v_i$ which is defined as 
\begin{equation}
\overline{X}^v_i = e^{\gamma^v_i}, i=1,2,\ldots,N
\label{eqs26}
\end{equation}
gives a scattering intensity (by means of the normalised matrix multiplication in Eq. \eqref{eqs25}) $\hat{I}_j$ which is closest to $\hat{I}^{\ast}_j$. The search for the parameters $\gamma^v_i$ is done by solving a weighted least square problem similar to that in Eq. (6) of the main text given as 
\begin{equation}
\min f_v = \sum_{j=1}^{M}{w_j^v\left[\hat{I}^{\ast}_j - \hat{I}_j\right]^2},
\label{eqs27}
\end{equation}
where $w_j^v$ is a weight function similar to the case of Eq. \eqref{eqs19} for the number based PSD.

The optimisation problem in Eq. \eqref{eqs27} is solved using the Levenberg-Marquardt algorithm as implemented in Matlab.
Since the scattering intensity $\overline{I}_j$ in Eq. \eqref{eqs18} was constructed from a number based PSD obtained close to the global minimum of the objective function $f$ in Eq. (6) of the main text, then an initial estimate $X^{v0}_i$ of the volume based PSD which is close to the global minimum of the objective function $f_v$ in Eq. \eqref{eqs27} can be constructed by the method of truncated singular value decomposition (TSVD) \cite{Aster2013} using the scattering intensity $\hat{I}_j$ in Eq. \eqref{eqs25}. This initially estimated volume based PSD $X^{v0}$ can then be used to obtain an initial starting solution for $\gamma^v_i$ and passed on to the Levenberg-Marquardt algorithm to solve the optimisation problem in Eq. \eqref{eqs27}. The TSVD method can only be used to generate an initial estimate for $\gamma_i^v$ as it  predicts PSDs with negative values because of the ill-conditioning of the inverse problem. However, the formulation of the volume based PSD $\overline{X}^v_i$ in Eq. \eqref{eqs26} guarantees non-negative values. 

Finally, the volume based PSD is normalised as
\begin{equation}
\hat{X}^v_i = \frac{\overline{X}^v_i}{\sum_i^N{\overline{X}^v_i}}.
\label{eqs28}
\end{equation}

\end{document}